\documentclass[pra,twocolumn,amsmath,amssymb,groupedaddress,superscriptaddress]{revtex4}
\usepackage{graphics,color}
\usepackage{amsmath}
\usepackage[dvips]{graphicx}
\usepackage{float,graphicx}
\usepackage{subfigure,hyperref}
\usepackage{epstopdf}
\usepackage{epstopdf}
\usepackage{soul}
\usepackage{epstopdf}
\newcommand{\beq}{\begin{equation}}
\newcommand{\eeq}{\end{equation}}
\newcommand{\beqa}{\begin{eqnarray}}

\newcommand{\eeqa}{\end{eqnarray}}
\newcommand{\unit}{1\!\!1}

\newcommand{\abs}[1]{\left\vert #1\right\vert}

\def\be{\begin{equation}}

\def\ee{\end{equation}}

\def\mmax{\textmd{max}}

\def\iint{\textmd{int}}

\def\cch{\textmd{ch}}

\def\ttot{\textmd{tot}}
\def\oopt{\textmd{opt}}

\def\be{\begin{equation}}
\def\ee{\end{equation}}
\def\bea{\begin{eqnarray}}
\def\eea{\end{eqnarray}}

\usepackage{bm}
\usepackage{graphicx}

\begin{document}
\title[Non-Markovianity and initial state memory]{Non-Markovianity and memory of the initial state}

\author{Margarida Hinarejos}
\affiliation{Instituto de F\'isica, Facultad de Ingenier\'ia, Universidad de la Rep\'ublica,  Av. Julio Herrera y Reissig 565,  11300 Montevideo, Uruguay}
\author{Mari-Carmen Ba\~nuls}
\affiliation{Max-Planck-Institut f\"ur Quantenoptik, Hans-Kopfermann-Str. 1, 85748 Garching, Germany.}
\author{Armando P\'erez}
\affiliation{Departament de F\'isica Te\`orica and IFIC, Universitat de Val\`encia-CSIC, Dr. Moliner 50, 46100-Burjassot, Spain}
\author{In{\'e}s de Vega}
\affiliation{Department of Physics and Arnold Sommerfeld Center for Theoretical Physics, Ludwig-Maximilians-University Munich, Germany}

\begin{abstract}
We explore in a rigorous manner the intuitive connection between the non-Markovianity of the evolution of an open quantum system and the performance of the system as a quantum memory. Using the paradigmatic case of a two-level open quantum system coupled to a bosonic bath, we compute the recovery fidelity, which measures the best possible performance of the system to store a qubit of information. We deduce that this quantity is connected, but not uniquely determined, by the non-Markovianity, for which we adopt the BLP measure proposed in \cite{breuer2009}. We illustrate our findings with explicit calculations for the case of a structured environment.
\end{abstract}
\maketitle

 \section{Introduction}
 
Open quantum systems present a very rich dynamics that is described by tracing out the environment degrees of freedom, either analytically or numerically, and thus considering the environment only through its action onto the system. 
A non-Markovian theory \cite{devega2015c}, which does not rely on assuming that the environment relaxes instantaneously when interacting with the system, appears to be necessary to describe systems such as superconducting flux-qubits coupled to waveguides or to complex environments \cite{xiang2013,peropadre2013}, processes in surface science \cite{baer1997,koch2003,asplund2011} and solutions \cite{koch2002,gelman2004}, or atomic emission in structured environments such as photonic crystals \cite{florescu2001,devega2005,devega2008c,devega2014a}. Besides that, inspired by the earlier works \cite{wolf2008,breuer2009,rivas2010a}, a large number of theoretical proposals have emerged to characterize the non-Markovian character of an evolution (see \cite{rivas2014,breuer2015,devega2015c} for reviews on the subject). The non-Markovianity, as quantified by such measures, has been determined to be a resource for quantum information tasks such as quantum communication \cite{laine2014,hengliu2013}, efficient superdense coding in the presence of dephasing noise \cite{liu2015},  entanglement generation \cite{braun2002,bellomo2007,paz2008,valido20131,valido20132}, quantum metrology \cite{matsuzaki2011,Chin2012a} and information transfer through a noisy channel \cite{ziman2002,maniscalco2007,bylicka2014,caruso2014}. 
 
Non-Markovianity (NM) holds an intuitive connection to memory. Based on this idea, it has been suggested that non-Markovianity can be exploited to engineer quantum memories \cite{rosario2015,man2015,man2015b}. 
Indeed, NM is linked to a back flow of information from the environment into the system, as evidenced by the Breuer-Laine-Piilo (BLP) NM measure \cite{breuer2009}.
However, to establish a rigorous connection between quantum memories and NM requires a quantitative comparison which, to the best of our knowledge, has not yet been performed. 
In this work, we analyze this question and show that there is not a direct relation between NM and the functioning of the system as a quantum memory.  

To this end, we consider the simplest, yet paradigmatic case, of a two level system coupled to a structured environment. The capability of the system to serve as a quantum memory can be quantified by the optimal recovery fidelity, which is a well-defined quantum information quantity \cite{bowdrey2002}. Indeed, we find that this quantity is linked not only to the NM, but also to information losses of the system, 
so that non-Markovianity is neither a sufficient nor a necessary condition to achieve a certain value of the optimal recovery fidelity. 
As will be illustrated here, there are situations in which the NM is large, and yet the system loses its information (i.e. it \textit{forgets} about its initial state) completely. And conversely, the NM may remain small while the system preserves memory of its initial state, for instance because of the presence of some symmetry.


The paper is organized as follows. In Sec. \ref{memory} we introduce our model, as well as the optimal recovery fidelity as a measure to estimate the memory of the system's initial state. We also establish a link between such optimal recovery fidelity and the NM of a process. In addition, several bounds relating these quantities to the system losses are also derived. Sec. \ref{pseudogap_model} illustrates these ideas by analyzing the example of a two level system coupled to a highly structured environment, characterized by a spectral density that can be tuned to display a pseudogap (i.e. a dip in its shape), and a gap (i.e. a region where it vanishes). 
We consider two different cases. In the first one, the rotating wave approximation (RWA) is applicable (Sec. \ref{RWA}), and the problem can be solved
exactly. As a second example we consider a spin-boson model (Sec. \ref{SB}) and analyze it numerically using matrix product states (MPS).

\section{Memory and non-Markovianity}

\label{memory}

The evolution of an open quantum system for a certain time, $t$, can be expressed
as the result of a quantum channel acting on an initial state
of the system, $\rho(t)=\phi_{t}[\rho_{s}(0)]$.
We will in the following focus on the case of a two level system, or a qubit.
The recovery fidelity quantifies the average fidelity with which 
a certain recovery operation, ${\mathcal{R}}_{t}$, 
can retrieve a pure initial state of the qubit \cite{bowdrey2002},
\begin{eqnarray}
F({\mathcal{R}}_{t})=\int d\mu_{\psi}\langle\psi|{\mathcal{R}}_{t}\phi_{t}[|\psi\rangle\langle\psi|]|\psi\rangle\label{eq:RF}.
\end{eqnarray}
To characterize the performance of the system as a quantum memory,
one can consider the optimal recovery fidelity over all possible recovery operations,
which is given by  \cite{mazza2013}
\begin{eqnarray}
F^{\textmd{opt}}(t)=\max_{{\mathcal{R}}_{t}}F({\mathcal{R}}_{t})=\frac{1}{2}+\frac{1}{6}\sum_{\alpha=x,y,z}{\mathcal{D}}_{\alpha}(t),\label{bound}
\end{eqnarray}

\begin{eqnarray}
{\mathcal{D}}_{\alpha}(t)=\frac{1}{2} \left \| \phi_{t}[|\alpha+\rangle\langle\alpha+|]-\phi_{t}[|\alpha-\rangle\langle\alpha-|] \right \| .
\label{distingu}
\end{eqnarray}
In the above expression, $\|\cdot\|$ is the trace norm, $\|A\|=\mathrm{tr}(\sqrt{AA^{\dagger}})$,
and $|\alpha\pm\rangle$ are the eigenstates of Pauli matrices, $\sigma_{\alpha}|\alpha\pm\rangle=\pm|\alpha\pm\rangle$ ($\alpha=x,\,y,\,z$).
The quantity ${\mathcal{D}}_{\alpha}(t)={\mathcal D}[\rho_{\alpha}^{+}(t),\rho_{\alpha}^{-}(t)]$ is, by definition,
the distinguishability of the states $\rho_{\alpha}^{+}(t)$ and $\rho_{\alpha}^{-}(t)$,
resulting from the evolution of the initially orthogonal states $|\alpha\pm\rangle$. Notice that, according to equation (\ref{bound}), $F^{\textmd{opt}}(t)\ge 1/2$. Therefore, the lower limit corresponds to a complete loss of memory of the initial state. 

In our case, the channel corresponds to 
a unitary evolution of system plus bath, with the environment initially in the vacuum state and using  the evolution operator ${\mathcal{U}}(t)=\exp(-iHt)$
for the total system Hamiltonian, $H$, 
followed by tracing out the environment.
Hence, we compute $\rho_{\alpha}^{\pm}=\phi_{t}[|\alpha\pm\rangle\langle\alpha\pm|]=\textmd{Tr}_{B}\{{\mathcal{U}}^{-1}(t)|\alpha\pm\rangle\langle\alpha\pm|{\mathcal{U}}(t)\}$,
where $\textmd{Tr}_{B}\{\cdots\}$ represents the trace over the environment.
We can always write
\begin{equation}
\rho_{\alpha}^{\pm}(t)=\phi_{t}[|\alpha\pm\rangle\langle \alpha\pm|]=\frac{1}{2}(\unit{+}\vec{P}_{\alpha}^{\pm}(t)\cdot\vec{\sigma}),
\end{equation}
where 
$\vec{P}_{\alpha}^{\pm}(t)$
is the polarization vector, with components $P_{\alpha,\alpha'}^{\pm}(t)$,
$\alpha'=x,y,z$. Then,
\begin{eqnarray}
\rho_{\alpha}^{+}(t)-\rho_{\alpha}^{-}(t)=\sum_{\alpha'=x,y,z}\frac{1}{2}\Delta_{\alpha,\alpha'}(t)\sigma_{\alpha'},
\end{eqnarray}
with $\Delta_{\alpha,\alpha'}(t)=P_{\alpha,\alpha'}^{+}(t)-P_{\alpha,\alpha'}^{-}(t)$.
One then obtains 
\begin{eqnarray}
 F^{\textmd{opt}}(t)=\frac{1}{2}+\frac{1}{12}\sum_{\alpha=x,y,z}\sqrt{\sum_{\alpha'}\Delta_{\alpha,\alpha'}^{2}(t)}.
 \label{optimalfin}
\end{eqnarray}
An alternative expression for $F^{\textmd{opt}}(t)$ can be obtained by recasting the action of the channel as a map of polarization vectors on the Bloch sphere \cite{hall2014}
\begin{equation}
\vec{P}(t)=M(t)\vec{P}(0)+\vec{q}(t).
\label{eq:polmap}
\end{equation}
We can write $\Delta_{\alpha,\alpha'}(t)=2M_{\alpha'\alpha}(t)$, so that $\sum_{\alpha=x,y,z}\sqrt{\sum_{\alpha'}\Delta_{\alpha,\alpha'}^{2}(t)}=2\|M(t)\|_{2,1}$, where $\|A\|_{2,1}\equiv\sum_{j}\sqrt{\sum_{i}|A_{i,j}|^{2}}$ is the $L_{2,1}$ norm of matrix $A$. The optimal fidelity in Eq. (\ref{optimalfin}) can thus be expressed as:
\begin{equation}
F^{\textmd{opt}}(t)=\frac{1}{2}+\frac{1}{6}\|M(t)\|_{2,1}.
\label{optimalfin_2}
\end{equation}

 In the following, we consider the measure of NM proposed by Breuer-Laine-Piilo (BLP measure) \cite{breuer2009}. In this proposal, a system is considered to be non-Markovian when there is a backflow of information from the environment to the system during the evolution. 
 This backflow of information is characterized by an increase in the distinguishability between pairs of evolving quantum states. 
More precisely, a system is non-Markovian if there is a pair of initial states $\rho_1(0)$ and $\rho_2(0)$, such that for certain time intervals at $t>0$ their distinguishability increases, 
\begin{eqnarray}
\sigma(\rho_1(0),\rho_2(0);t)=\frac{d}{dt}{\mathcal D}[\rho_1(t),\rho_2(t)]>0.
\end{eqnarray}
Following the BLP criterion, the amount of NM of a quantum process in a time interval (0,t) can be quantified as 
\begin{eqnarray}
N(t):=
\max_{\rho_1,\,\rho_2}
\int^t_{0,\sigma>0} ds \ \sigma(\rho_1(0),\rho_2(0),s).
\label{nonBreuer}
\end{eqnarray}
The maximization is over all possible pairs of initial states, $\rho_{1,2}(0)$.
Thus, $N$ reflects the maximum amount of information that can flow back to the system for a given process within the interval $(0,t)$. Notice that when the time interval is $(0,\infty)$, Eq. (\ref{nonBreuer}) corresponds to the definition of the BLP NM measure.
We may now write the optimal recovery fidelity, (\ref{bound}), for the channel corresponding to this evolution as
\begin{eqnarray}
F^\oopt(t)=1+\frac{1}{6}\sum_{\alpha=x,y,z}\left({\mathcal N}_\alpha(t)+{\mathcal P}_\alpha(t)\right).
\label{bound_NM}
\end{eqnarray}
where we have considered that $F^\oopt({t=0})=1$, and defined the quantities
\begin{eqnarray}
&{\mathcal N}_\alpha(t)&=\int^t_{0,\sigma>0}ds\  \sigma(\rho_{\alpha+}(0),\rho_{\alpha-}(0),s),\cr
&{\mathcal P}_\alpha(t)&=\int^t_{0,\sigma<0}ds\  \sigma(\rho_{\alpha+}(0),\rho_{\alpha-}(0),s).
\label{Losses}
\end{eqnarray}
The accumulated memory gains are then $\sum_{\alpha=x,y,z}{\mathcal N}_\alpha(t)$, and the memory losses, $\sum_{\alpha=x,y,z}{\mathcal P}_\alpha(t)$.
 In the above formulas, we have considered three different pairs of initial states, $\{|\alpha+\rangle,|\alpha-\rangle\}$, for three arbitrary orthogonal directions $\alpha=x,y,z$, such that $\rho_{\alpha\pm}(0)=|\alpha\pm\rangle \langle\alpha\pm|$. 
Although Eq. (\ref{nonBreuer}) is maximized by a pair of initially orthogonal states \cite{wissmann2012}, the optimal direction does not need to coincide
 with the ones considered in (\ref{bound_NM}). In the most general case, it can be even time-dependent.
 Thus, in general each of the ${\mathcal N}_\alpha(t)$ represents a lower bound to the NM of the system.


Nevertheless, since equation (\ref{bound_NM}) holds for any set $x,y,z$ of mutually orthogonal directions, we can fix the $z$ axis along the direction, $\hat{z}$, in the Bloch sphere that optimizes the gains, such that ${\mathcal N}_{\hat{z}}(t)=N(t)$, with $N(t)$ the NM given by (\ref{nonBreuer}). Thus, we can rewrite (\ref{bound_NM}) as
 \begin{equation}
 F(t)\leq F^{\oopt}(t)= 1 +\frac{1}{6}N(t)+\frac{1}{6}\sum_{\alpha=\hat{x},\hat{y}}{\mathcal N}_\alpha(t)+\frac{1}{6}\sum_{\alpha=\hat{x},\hat{y},\hat{z}}{\mathcal P}_\alpha(t). 
 \label{Fopt}
\end{equation} 

This expression, the main result of the paper, shows that the NM is not the only significant factor to determine the system's memory of its initial state, as characterized by the optimal recovery fidelity.  Rather, the latter is the result of a balance between the NM (settled by the optimal direction $\hat{z}$), 
the memory gains from initial pairs along two directions, $\hat{x}$ and $\hat{y}$,  orthogonal to $\hat{z}$, and the total losses along the three directions, $\sum_{\alpha=\hat{x},\hat{y},\hat{z}}{\mathcal P}_\alpha(t)$, as defined in Eq. (\ref{Losses}). 

Using the properties of $F(t)$ and $N(t)$ we can establish some relations among the different terms in (\ref{Fopt}).
Since necessarily $F(t)\leq 1$, we have that 
\begin{equation}
N(t)\leq \abs{\sum_{\alpha=\hat{x},\hat{y},\hat{z}}{\mathcal P}_\alpha(t)} -\sum_{\alpha=\hat{x},\hat{y}}{\mathcal N}_\alpha(t) \label{bound4}
\end{equation}
and therefore
\begin{equation}
N(t)\leq \abs{\sum_{\alpha=\hat{x},\hat{y},\hat{z}}{\mathcal P}_\alpha(t)}, \label{bound5}
\end{equation}
Thus the NM is upper-bounded by the absolute value of the losses in the 
orthogonal set of directions that includes that of the optimal pair, $\hat{z}$.
 In addition, from (\ref{Fopt}) we find that 
\begin{equation}
 \abs{\sum_{\alpha=x,y,z}{\mathcal P}_\alpha(t)}\geq \sum_{\alpha=x,y,z}{\mathcal N}_\alpha(t)
 \label{bound17}
 \end{equation}
for any orthogonal set, $\{x,\,y,\,z\}$.
 Therefore, if a certain amount of information is "recovered", an even larger amount of information is necessarily lost during the same interval.
Finally, making use of the fact that ${\mathcal N}_\alpha(t)\leq N(t)$ for any direction, $\alpha$, we can upper-bound the optimal fidelity as
\begin{equation}
F(t)\leq F^{\oopt}(t)\leq 1 +\frac{1}{2}N(t)+\frac{1}{6}\sum_{\alpha=x,y,z}{\mathcal P}_\alpha(t).
\label{bound19}
\end{equation}

In the next section, we use several examples to illustrate the relations derived above among 
memory gains and losses and recovery fidelity, and explicitly show that NM does not necessarily imply
that the system retains memory of its initial state at the end of the evolution.

\begin{figure}[t]
\centering{}\includegraphics[width=0.95\linewidth]{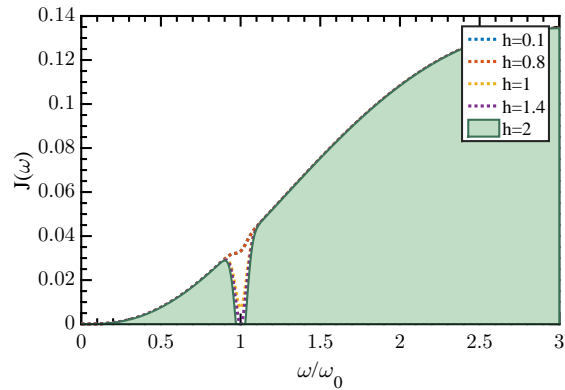}\caption{(Color online) Spectral density (\ref{spec}) for different values of the gap depth,
$h$. Other parameters are $\gamma=0.1$, $\omega_{0}=1$, $\eta=0.05\omega_{s}$,
and $\omega_{\textmd{max}}=3$. \label{spectral}}
\end{figure}

\section{A qubit in a structured environment}

\label{pseudogap_model}


Our basic model is the \emph{spin-boson} Hamiltonian \cite{leggett1987,devega2015c}, that describes a two-level quantum system coupled to a bosonic reservoir, 
\begin{eqnarray}
H & = & H_{\textmd{sys}}+\sum_{k}\,\tilde{g}(k)\,L(b_{k}+b_{k}^{\dagger})+\sum_{k}\,\omega(k)b_{k}^{\dagger}b_{k},\label{Hamil3_0}
\end{eqnarray}
where $H_{\mathrm{sys}}$ is the Hamiltonian of the two-level system,
the coupling operator is $L=\sigma_x=\sigma^++\sigma^-$, 
$\tilde{g}(k)$ are the coupling strengths, and $b_{k}$ ($b_{k}^{\dagger}$)
are the operators that annihilate (create) a harmonic mode of frequency $\omega(k)$, 
which satisfy canonical commutation relations, $[b_{k},b_{k'}^{\dagger}]=\delta_{k,k'}$.
The index $k$ labels the momentum of the modes, for which we assume a cutoff $k_{\textmd{max}}$. In the
frequency representation, and provided that the environment is initially
in a Gaussian state, this Hamiltonian can be rewritten as \cite{devega2015}
\begin{eqnarray}
H & = & H_{\textmd{sys}}+\sigma_{x}\sum_{\omega}g(\omega)\,(b_{\omega}+b_{\omega}^{\dagger})+\sum_{\omega}\,\omega b_{\omega}^{\dagger}b_{\omega},\label{Hamil3}
\end{eqnarray}
where we have defined $g(\omega)=\sqrt{J(\omega)}$, being $J(\omega)=\tilde{g}^{2}(\omega)\rho_{\textmd{DOS}}(\omega)$
the spectral density of the environment, corresponding to a density of states of the environment $\rho_{\textmd{DOS}}(\omega)$.
There will be a frequency cutoff, $\omega_{\textmd{max}}$, determined by $k_{\textmd{max}}$.

We analyze the case of an environment with a spectral density that can be tuned from presenting a small pseudogap (thus being slightly structured) to display a full gap (thus being highly structured and leading to non-Markovian dynamics). 
This type of spectral densities can be encountered for instance in photonic band gap materials, artificially generated materials that present a periodicity in the refractive index \cite{john1987,yablonovitch1987}. In this regard, either a pseudogap or a gap
can be obtained by simply varying the contrast between the refractive index of the periodic elements and the background material. Following the model in \cite{vats2002}, the density of states of the radiation field in these materials has the form
\begin{eqnarray}
J(\omega)=\eta\frac{\omega^2}{c^3}\bigg[1-h e^{-\big(\frac{\omega-\omega_0}{\eta}\big)^2}\bigg],
\label{spec}
\end{eqnarray}
with $h$ a dimensionless parameter describing the depth of the pseudogap, $\omega_0$ its central frequency, which we take to be $\omega_0=1$,
and $\eta$ its width. 
 Fig. \ref{spectral} represents the spectral density for various values of $h$. 

Specifically, we consider the system Hamiltonian $H_{\textmd{sys}}=\omega_s\sigma^+\sigma^-$, 
and assume that the characteristic frequency of the two-level system is
in resonance with the central frequency of the pseudogap, $\omega_s=\omega_0=1$.
Throughout all this section we take for the parameters of the bath the fixed values $\eta=0.05 \omega_{0}$, $\omega_{\mathrm{max}}=3\omega_0$.
 The parameter $h$ determines whether the gap in the spectral density is fully opened ($h>1$) or not ($h<1$).
Thus we select representative values within each range of values, namely $h=0.1$ and $h=1.4$.
 We consider two different regimes of the model, and analyze in each of them the behavior of memory and NM measures.


\subsection{An exactly solvable case} 

\label{RWA}


Under certain conditions \cite{devega2015c} the model (\ref{Hamil3}), common in a light matter interaction scenario, can be further simplified by assuming the rotating wave approximation to discard \textit{fast rotating} terms of the form $b^{\dagger}_\omega\sigma^{+}$, and $b_\omega\sigma^{-}$,
so that the Hamiltonian becomes
\begin{eqnarray}
H & = & \omega_s\sigma^+\sigma^-+\sum_{\omega}g(\omega)\,(b_{\omega}\sigma^{+}+b_{\omega}^{\dagger}\sigma^{-})+\sum_{\omega}\,\omega b_{\omega}^{\dagger}b_{\omega}.\cr
&&\label{Hamil4}
\end{eqnarray}
This model conserves the total number of excitations, what simplifies the problem considerably, 
 in particular when the environment is at zero
temperature. In this case, an exact master equation can be derived
\cite{vacchini2010,breuerbook,devega2015c}, which depends on time-dependent dissipation
rates. Its solution can be written as
\begin{eqnarray}
\rho(t)&=&\left(\begin{array}{cc}
|\Gamma(t)|^{2}\rho_{++}(0) & \Gamma(t)\rho_{+-}(0)\\
\Gamma(t)^{*}\rho_{-+}(0) & 1-|\Gamma(t)|^{2}\rho_{++}(0)
\end{array}\right), \label{eq:twolelevelrho}
\end{eqnarray}
where $\rho_{ij}(t)$ are the components of the reduced density operator
in the $\sigma_{z}$ eigenbasis, and $\Gamma(t)$ is
the solution of the integro-differential equation
\begin{equation}
\frac{d}{dt}\Gamma(t)=-\int_{0}^{t}dt_{1}f(t-t_{1})\Gamma(t_{1}),
\label{eq:integrodiff}
\end{equation}
with initial condition $\Gamma(0)=1$, being $f(t)$  the two-point
correlation function of the environment, $f(t-t_1)=\sum_kg_k^2e^{i(\omega_s-\omega_k)(t-t_1)}$.



It is easy to see that the optimal fidelity in Eq. (\ref{optimalfin_2}) depends only on the instantaneous value $|\Gamma(t)|$, 
\begin{equation}
F^{\textmd{opt}}(t)=\frac{1}{2}+\frac{1}{6}(|\Gamma(t)|^{2}+2|\Gamma(t)|).\label{eq:FoptGamma}
\end{equation}

Although there is no simple analytical solution to Eq. (\ref{eq:integrodiff}) for the spectral density (\ref{spec}), 
the exact time dependence (\ref{eq:twolelevelrho}) can be found numerically.
This allows us to analyze in detail the behavior during the evolution of the
memory gains and losses, the optimal fidelity and the non-Markovianity.

\begin{figure}[t]
\centering{}\includegraphics[width=1\linewidth]{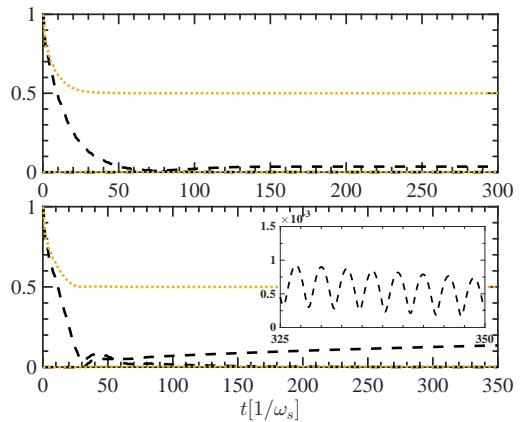}\caption{(Color online) RWA case: 
Evolution of the distinguishabilities ${\mathcal D}_y$  and ${\mathcal D}_z$ from their initial value of one, in black-dashed and orange-dotted lines respectively. We also represent the evolution of the NM lower bounds ${\mathcal N}_\alpha$ with $\alpha=y,z$, from their initial value of zero, corresponding to black-dashed and orange-dotted lines respectively ($x$ and $y$ components are indistinguishable in both cases).
Pseudogap depths $h=0.1$ (top panel); and $h=1.4$ (bottom panel). 
The inset represents a detail of the distinguishability components ${\mathcal D}_{x,y}$ for $h=1.4$. The observed oscillations lead to a growing ${\mathcal N}_x$. 
\label{Fig1RWA}}
\end{figure}

\begin{figure}[t]
\centering{}\includegraphics[width=1\linewidth]{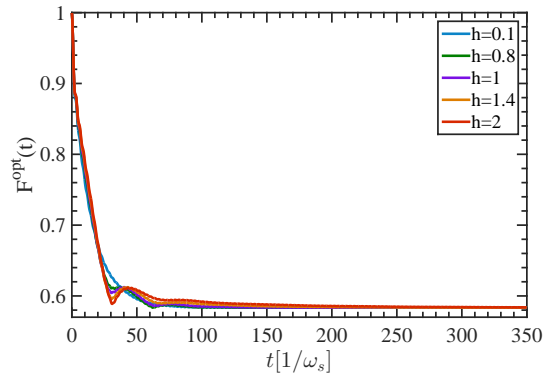}\caption{(Color online) RWA case: 
Evolution of the optimal fidelity for several values of $h$. Faster decaying curves correspond to smaller values of $h$.
\label{gainlosses_RWA}}
\end{figure}

\begin{figure}[t]
\centering{}\includegraphics[width=1\linewidth]{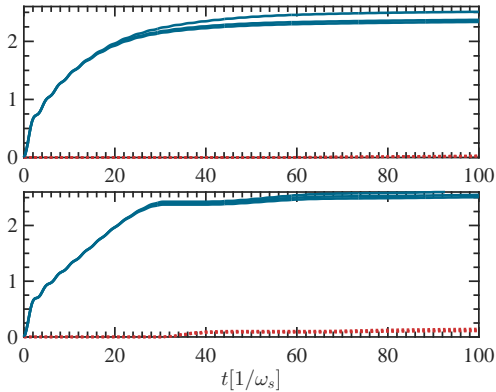}\caption{(Color online) RWA case: Evolution of the losses (blue dotted) and gains (red solid lines) in the process, corresponding respectively to the l.h.s. and r.h.s. of bound (\ref{bound17}). The curves correspond to three typical values for the angle $\theta$ between $0$.
We also consider two different values of the pseudogap depth $h=0.1$ (top panel) and $h=1.4$ (bottom panel).
\label{Fbound17}}
\end{figure}


As discussed in Sec. \ref{memory}, the accumulated memory gains corresponding to any direction, ${\cal N}_{\alpha}(t)$, set a lower bound for the NM.
We start by considering, for two different depths of the pseudogap, the bounds given by the quantization directions, $x$ and $z$ 
(since the model is invariant under rotations of the two-level system 
about the $z$ axis, directions $x$ and $y$ are identical 
\footnote{This invariance is a remainder of the conservation of total number of excitations, $\frac{1+\sigma_z}{2} + \sum_k b_k^{\dagger}b_k$, 
when we consider a fixed state of the environment to derive the master equation.})
 and compare them to the corresponding time dependent distinguishabilities, 
${\cal D}_{\alpha}(t)$, which determine the optimal recovery fidelity, (\ref{bound}).
The results, in Fig. \ref{Fig1RWA}, show that in the fully gapped case, $h=1.4$, the ${\mathcal{N}}_{x}(t)$ bound grows steadily in time, 
indicating that the system keeps regaining memory along the evolution. 
Instead, when we compute the optimal recovery fidelity, (\ref{bound}), which is shown in Fig. \ref{gainlosses_RWA}, we 
find that it reaches the stationary value $~0.58$,
relatively close to the minimal one, $0.5$, meaning that there is little memory of the initial state.
This apparent contradiction is explained by looking at the distinguishabilities, ${\cal D}_{\alpha}(t)$. Indeed
(see inset of bottom panel in Fig. \ref{Fig1RWA}), 
although the distinguishability ${\cal D}_{x}(t)$ decreases on average, it shows persistent oscillations, such that
during each period there will be some \emph{information} gained, which causes the accumulated
${\mathcal{N}}_{x}(t)$ to grow in time.
Corresponding to the gain, nevertheless, there is also a loss of information during each oscillation, and thus
 ${\mathcal P}_x(t)$ in Eq. (\ref{Losses}) also grows in time. Moreover,  ${\mathcal P}_x(t)$ gets larger
than ${\cal N}_{x}(t)$ in absolute value, since the average value 
of ${\cal D}_{x}(t)$ is decreasing.

However, the behavior of the $z$ component is rather different. We observe that at long times ${\mathcal D}_z(t)$ appears to converge to $1/2$,
 while the corresponding gain, $\mathcal{N}_z(t)$, remains constant and much smaller than ${\mathcal N}_{x}(t)$. 
In this case, the non-vanishing distinguishability indicates that the system conserves some memory of the initial state. This is
however not directly related to information gains associated to NM, but it responds to the symmetry (conservation of 
the number of excitations) present in the problem. 


According to the bound (\ref{bound17}), the combined information gains, $\sum_{\alpha=x,y,z}{\mathcal N}_{\alpha}(t)$,
are upper bounded by the absolute value of the combined information losses. 
Since we have an exact solution of the problem, we can compute the gains and losses,
${\mathcal N}_{(\theta,\phi)}(t)$, ${\mathcal P}_{(\theta,\phi)}(t)$, associated to any direction 
on the Bloch sphere, $\hat{n}(\theta,\phi)=(\sin{\theta}\cos{\phi},\sin{\theta}\sin{\phi},\cos{\theta})$.
Thus, we can analyze the bound (\ref{bound17}) for an arbitrary set of mutually orthogonal directions, $\{x',\,y',\,z'\}$,
specified by setting the $z'$ direction $\hat{z'}=\hat{n}(\theta,\phi)$.
Fig. \ref{Fbound17} confirms that the bound is not tight, whatever the chosen set of directions, 
and that, at least in the present example, the losses, rather than the gains, are 
the most important factor to determine the optimal recovery fidelity and thus the memory of the system's initial state. 



\begin{figure}[t]
\centering{}\includegraphics[width=1\linewidth]{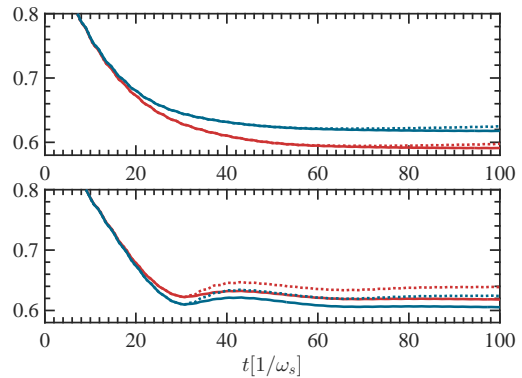}\caption{(Color online) RWA case: Evolution of the right-hand side and left-hand side of the bound (\ref{bound19}), in dotted and solid lines respectively, for two typical values for the angle $\theta$ ($\theta=0.07$ and $\theta=0.032$ corresponding to red and blue colors respectively), and considering two different values of the pseudogap depth, $h=0.1$ (top panel) and $h=1.4$ (bottom panel). \label{Fbound19}}
\end{figure}

Eq. (\ref{bound19}) provides now a bound to the optimal recovery fidelity in terms of NM and total information losses for an arbitrary set of directions. To analyze this bound, we shall consider that the NM measure, $N(t)$, corresponds to the maximum of ${\mathcal{N}}_{(\theta,\phi)}(t)$ over all angles, as Eq. (\ref{nonBreuer}) is maximized by a pair of fully polarized initial states, 
$\rho_{\hat{n}\pm}(0)=|\hat{n}\pm\rangle\langle\hat{n}\pm|$ \cite{wissmann2012}. 
Given the symmetry of the model under $z$ rotations, we can ignore the azimuthal angle, and 
we find numerically that the maximum is attained for $\theta=\pi/2$.
We can compare now both sides of the inequality (\ref{bound19}) for different choices of the set of directions, and find that  
the bound is relatively tight in all cases, for both pseudogap depths considered, $h=0.1$ and $h=1.4$,
as illustrated in Fig. \ref{Fbound19}.


For the direction of maximum gains, $\theta=\pi/2$, the derivative of the distinguishability
can be written $\sigma\left(|\hat{n}+\rangle,|\hat{n}-\rangle,t\right)=d|\Gamma(t)|/dt$,
and thus the NM (\ref{nonBreuer}) reads
\begin{equation}
N(t)=\sum_{i}(|\Gamma(t_{i+1})|-|\Gamma(t_{i})|),\label{eq:NMGamma}
\end{equation}
where the sum is over all time intervals $[t_{i},t_{i+1}]$, for $0\leq t_{i,\, i+1}\leq t$, which satisfy
 $\left.\frac{d |\Gamma(t)|}{d t}\right |_{\tau}>0$  for $\tau\in[t_i,t_{i+1}]$.

This allows us to see that 
there is a relation between the \textit{instantaneous} recovering of the memory of the initial state, and the appearance of a backflow of information from the environment. 
Indeed, the derivative of Eq. (\ref{eq:FoptGamma}) shows that $d F^{\textmd{opt}}(t)/dt\leq 0$ if and only if $d |\Gamma(t)|/dt>0$, 
i.e. when there is a backflow of information from the environment into the system.
However, there is no link between the optimal recovery fidelity at a certain time $t$ and the total amount of backflow that has occurred up to this time. In this regard, we observe from Eqs. (\ref{eq:FoptGamma}) and (\ref{eq:NMGamma}) that while the optimal recovery fidelity at time $t$ is obtained from the instantaneous value of $|\Gamma(t)|$, the value of the NM
measure at this time depends on the full history of the derivative
$\frac{d}{dt}|\Gamma(t)|$ during the interval $(0,t)$. It is therefore expected
that, in general, both quantities will behave in an independent way.

Finally, notice that the condition for backflow of information, $d |\Gamma(t)|/dt>0$, is indeed related
to the appearance of negative damping rates in the master equation that governs the dynamics of the system \cite{hall2014}.
From the map $\rho(t)=\phi_t[\rho(0)]$ we can write a master equation,
$d\rho(t)/dt={\mathcal L}(t)\rho(t)$, with ${\mathcal L}(t)=\frac{d\phi_t}{dt}\phi_t^{-1}$, provided that 
the inverse map $\phi_t^{-1}$ exists.
From the map defined by Eq. (\ref{eq:twolelevelrho}) we find for our problem
\bea
{\mathcal L(t)}\rho(t)&=&-\frac{i}{2}S(t)[\sigma^+\sigma^-,\rho(t)]\cr
&+&\gamma(t)[\sigma^-\rho\sigma^+-\frac{1}{2}\{\sigma^+\sigma^-,\rho(t)\}],
\eea
where $S(t)$, a time-dependent Lamb shift, and $\gamma(t)$, the damping rate, are defined as
\bea
S(t)&=&-2\mathrm{Im}\left\{\frac{1}{\Gamma(t)}\frac{d}{dt}\Gamma(t)\right\},\cr
\gamma(t)&=&-2\mathrm{Re}\left\{\frac{1}{\Gamma(t)}\frac{d}{dt}\Gamma(t)\right\}=-\frac{2}{|\Gamma(t)|}\frac{d}{dt}|\Gamma(t)|,
\eea
and the generator ${\mathcal L(t)}$ will be well-defined as long as $\Gamma(t)\neq 0$. 
This shows that, indeed, the condition $d |\Gamma(t)|/dt>0$ is equivalent to $\gamma(t)<0$.

\subsection{More general scenario} 
\label{SB}

In some cases the RWA is not be applicable to the model (\ref{Hamil3}) and the total number of excitations is not conserved.
To study the dynamics of the relevant quantities we thus treat the full problem numerically,
using matrix product state (MPS) techniques \cite{vidal2003,verstraete2008,scholl2011,scholl2005}. 
Although the focus of this paper is not on physical realizations, we notice here that such full model may be of relevance to describe experimental settings that achieve an ultra-strong coupling regime, such as superconducting circuits \cite{niemczyk2010}, superconducting qubits in open transmission lines \cite{peropadre2013}, coupled-cavity polaritons \cite{gunter2009}, or plasmon polaritons in semiconductor quantum wells \cite{geiser2012} (see also \cite{devega2015c} for a discussion on the subject). 



Unlike the RWA regime discussed in the previous section, the general case does not admit an exact solution and the numerical results are only approximate. 
In our formalism, the errors come from three different truncations.
First of all, in order to deal with the dynamics of the full system including the environment, we use a 
 representation of the environment \cite{prior2010,devega2015} (see also Appendix A for details), which maps the bath to
a semi-infinite bosonic chain. In practice, however, we need to work with a finite chain, and thus truncate the number of  
environmental modes included in the evolution. 
A second error source is the truncation of the maximum occupation number of the environmental bosonic modes, in order
to have finite dimensional local Hilbert spaces, as required by the MPS formalism.
 Finally, the MPS ansatz used to describe the state of the whole system, has a finite maximal bond dimension, $D$,
 determining the size of the tensors that compose the ansatz, and the precision of the approximation.
 By repeating our simulations for varying values or all three truncation parameters, we can estimate the 
effect of each truncation and therefore ensure the reliability of our results.

\begin{figure}[t]
\centering{}\includegraphics[width=1\linewidth]{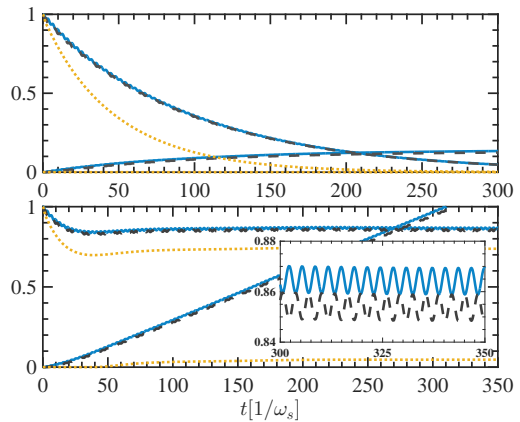}
\caption{(Color online) Full model:
Evolution of the distinguishabilities from their initial value of one, the components ${\mathcal D}_y$, ${\mathcal D}_x$, and ${\mathcal D}_z$ in blue-solid, black-dashed and orange-dotted lines respectively. We also represent the NM lower bound from the initial value of zero, the components ${\mathcal N}_\alpha$ for $\alpha=x,y,z$ corresponding again to blue-solid, black-dashed and orange-dotted lines respectively.
We have considered pseudogap depths $h=0.1$ (upper) and $1.4$ (lower pannel).
The inset in the lower plot  shows the oscillations of ${\mathcal D}_{x,y}(t)$ for $h=1.4$, responsible for the linear growth 
of the corresponding ${\mathcal N}_{x,y}(t)$ with time.
Results were obtained with MPS of bond dimension $D=60$. 
\label{Fig3}}
\end{figure}


Following the analysis of the previous section, we start by studying the lower bounds to NM imposed by the information gains that correspond to fully polarized initial states along the quantization axes.
As illustrated in the upper panel of Fig. \ref{Fig3} for $h=0.1$, we find that all three distinguishabilities seem to decay to zero, meaning that the system
completely loses memory of its initial state, as initially orthogonal pairs become indistinguishable.
In contrast to the RWA regime, now there is no conserved quantity that protects some component.
Thus in the long time limit, the optimal fidelity will decay to its minimal value, $1/2$, meaning that the best protocol for recovery will not do better than a random guess.

\begin{figure}[t]
\centerline{\includegraphics[width=0.95\linewidth]{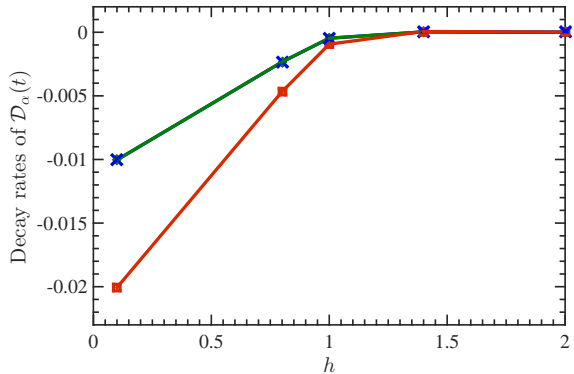}}
\caption{(Color online) Full model: Decay rates of the distinguishabilities ${\mathcal D}_\alpha(t)$ for $\alpha=x,y,z$ (blue crosses, red squares and green diamonds respectively) as a function of the gap depth, $h$.
 \label{rates}}
\end{figure}

Nevertheless, the lower bound of the NM saturates to a non-vanishing value, which we find to be larger as we increase $h$, if we stay within the not-fully-gapped regime, $h<1$.
The decay of the distinguishabilities in these cases can be fitted by an exponential, with a decay rate that decreases for larger $h$, as shown in  Fig. \ref{rates}, as we approach 
the limiting case, $h=1$, for which the gap starts to open.
In such limiting case the distinguishabilities decay, but very slowly. Under the assumption that this behavior would persist at longer times the system will lose its memory after a long time, while the lower bound of NM saturates to a certain value as soon as ${\mathcal D}_{\alpha}$ become zero. In summary, the results for $h\le1$ show that a finite and possibly large value for the non-Markovianity in the long time limit does not imply a finite value for the optimal recovery fidelity, which is determined by the sum of distinguishabilities.

For $h>1$, as exemplified by $h=1.4$ in the lower panel of Fig. \ref{Fig3}, the distinguishabilities do not longer decay, but approach to relatively large asymptotic values. Nevertheless, 
 their detailed evolution shows that their value exhibits persistent oscillations 
corresponding to a sustained exchange of information between the system and the environment.
As discussed in the previous section, this implies a lower bound of NM which grows monotonically in time, as can be appreciated in the figure. Thus, the NM will diverge as $t\to\infty$. In conclusion, also in this case, the amount of the non-Markovianity does not predict the magnitude of the system's memory of the initial state.


We note that in the latter case, the BLP NM measure
 can be modified to avoid the divergence, as it has been done in \cite{rivas2014} for the Rivas-Huelga-Plenio (RHP) measure. 
 Following a similar procedure, 
 we could also define a finite-valued modified BLP NM measure.
 However, its value would still not directly determine the memory of the system.

\begin{figure}[t]
\centering{}\includegraphics[width=1\linewidth]{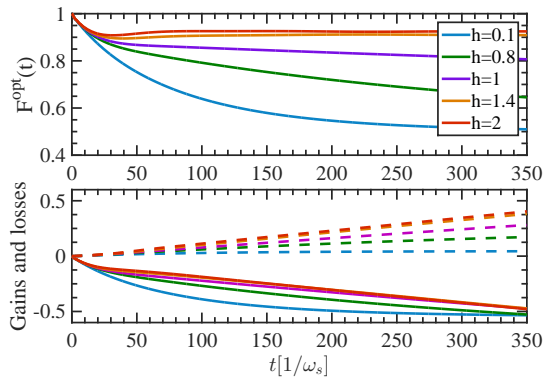}\caption{(Color online) Full model: Evolution of the optimal fidelity (upper panel) and total memory gains, $\frac{1}{6}({\mathcal N}_x+{\mathcal N}_y+{\mathcal N}_z)$  (dotted) and the losses  $\frac{1}{6}({\mathcal P}_x+{\mathcal P}_y+{\mathcal P}_z)$ (solid line), for different values of $h$ (faster decaying and slower growing curves correspond to smaller values of $h$).
 \label{gainlosses_noRWA}}
\end{figure}

Finally, we can also compare the total accumulated memory gains, $\sum_\alpha {\mathcal N}_\alpha$, and the total amount of losses $\sum_\alpha {\mathcal P}_\alpha$, 
and see how they combine in the optimal fidelity recovery, according to Eq. (\ref{bound_NM}).
Figure \ref{gainlosses_noRWA} shows these quantities, and illustrates the idea
 that the optimal recovery fidelity is a balance between both gains and losses.
 In order to have a final fidelity larger than $1/2$, and thus some possible use of the system as a quantum memory,
 the accumulated gains need to overcome the total losses, as is the case for $h>1$.


\section{Conclusions}

In this paper we have rigorously analyzed the relation
between the potential use of an open two level system as a quantum memory,
as quantified by the optimal recovery fidelity, 
and the non-Markovianity of its evolution. 
Although naively NM is expected to result in the system maintaining some memory of its initial state, we have shown
that the relation between them
 crucially includes a quantity that measures the information losses in the system, produced by the dissipation. 


We have illustrated this result by analyzing the case in which the two level system
 is coupled to an environment which can be tuned from being slightly to highly structured. Indeed, by varying a parameter $h$ it
 may display a pseudogap (for $h<1$) or a full gap or vanishing region (for $h>1$).
Considering two different parameter regimes, we have shown that for a full gap in the spectral density of the environment, 
the non-Markovianity grows unboundedly in time, and thus cannot be 
a good quantifier of the magnitude of the memory of the initial state, which instead converges to a finite value.



The reason the NM is not a good indicator of a prevailing memory of the initial state is that it is determined by accumulated gains of information along the evolution, without regard for corresponding (and possibly larger) losses. Hence, 
the presence of NM alone is no guarantee for such a memory. In fact, in Sect. \ref{RWA} we have illustrated a regime of parameters in which the optimal fidelity attains a constant value, while the non-Markovianity
depends on the characteristics of the bath.


In conclusion, and at least under the conditions hereby considered, our analysis shows that neither the value nor the qualitative behavior of the non-Markovianity are good predictors of the long time memory of the initial state.

\textit{Acknowledgments}
The authors gratefully acknowledge A. Rivas for interesting discussions. This work has been supported by the Spanish Ministerio de Educaci\'on e Innovaci\'on, MICIN-FEDER projects  FPA2014-54459-P and SEV-2014-0398, and “Generalitat Valenciana” grant GVPROMETEOII2014-087. I.D.V has been financially supported by the Nanosystems Initiative Munich (NIM) (project No. 862050-2) and partially from the Spanish MINECO through project FIS2013-41352-P and COST Action MP1209. 
MH acknowledges financial support from ANII (Uruguay) grant PD-NAC-2014-1-102359. 

 \section{Appendix A: Chain representation}
 Let us consider the Hamiltonian (\ref{Hamil3}) in the continuum limit, 
\begin{eqnarray}
H&=&H_S+\int_{0}^{1}dkg(k)(b(k)L^\dagger+L b(k)^{\dagger})\cr
&+&\int_{0}^{1}dk\omega(k)b(k)^{\dagger}b(k),
\end{eqnarray}
with $b(k)$ $(b^\dagger(k))$ the continuous counterpart of $b_k$ $(b_k^\dagger)$, $g(k)$ is the continuous counterpart of the coupling strength $g_k$, and $\omega(k)$ the continuous counterpart of the dispersion $\omega_k$. In addition, we have rescaled the integrals, such that $\omega(1)=\omega_\mmax$, \textit{i.e.} the frequency cutoff of the environment. When the environment is in a Gaussian state, $\omega(k)$ and $g(k)$ enter the description of the OQS only through the spectral density, $J(\omega(k))=g^2(k)D(\omega(k))$, where $D(\omega(k))=|\partial\omega(k)/\partial k|^{-1}$ is the photonic density of states (DOS). 

One can reproduce the same spectral density by introducing a new dispersion relation, $\hat{\omega}(k)=\omega_c k$ (with $\omega_c$ an arbitrary constant that may be taken as one), such that $D(\hat{\omega(k)})=\omega_c$, and a new coupling $\hat{g}(k)$, such that $\hat{g}(k)=\sqrt{J(\omega(k))}$.
In terms of these new quantities, the continuum representation of the above Hamiltonian reads  
\begin{align}
\tilde{H}_{\ttot}&=H_S +\int_0^1 dk k a_{k}^\dagger a_{k}+\int_0^1 dk  \hat{g}(k) (L^\dagger a_{k}+a^\dagger_{k}L)
\label{h4}
\end{align}
Then, using the unitary transformation discussed in \cite{prior2010,chin2010b,devega2015}, new bosonic operators $B_n$ and $C_n$ can be defined for each reservoir, such that
\begin{eqnarray}
a_{k}=\sum_n U_{n}(k)B_n, 
\label{eq:BCn}
\end{eqnarray}
where $U_{n}(k)=g_j(k)\pi_{n}(k)/\rho_{n}$. Here, $\pi_{n}(k)$ are monic orthogonal polynomials that obey 
\begin{eqnarray}
\int_{0}^{1} dk J(k)\pi_{n}(k)\pi_{m}(k)=\rho_{n}^2\delta_{nm},
\end{eqnarray}
with $\rho^2_{n}=\int_{0}^{1} dk J(k)\pi^2_{n}(k)$ \cite{chin2010b,chinbook2011}. 
Hence, the proposed transformation is also orthogonal, $\int dk U^*_{n}U_{m}=\delta_{nm}$. The transformed Hamiltonian can be written as
$\tilde{H}^{\cch}_{tot}=H_S +H^{\cch}_B+\tilde{H}^{\cch}_\iint$,
with the interaction of the system with the first harmonic oscillator of each chain given by 
\begin{eqnarray}
\tilde{H}^{\cch}_\iint=g(L^\dagger B_0+B_0^\dagger L),
\end{eqnarray}
where $g=\rho_{0}$, and the Hamiltonian of the two chains given by 
\begin{align}
\tilde{H}^{\cch}_B=&\sum_{n=0,\cdots,M}(\alpha_n B_n^\dagger B_n+\sqrt{\beta_{1,n+1}}B^\dagger_{n+1}B_n+h.c.)
\end{align}
In order to perform the mapping, the recurrence relation of the orthogonal polynomials have been used, namely 
\begin{eqnarray}
\pi_{n+1}(k)=(k-\alpha_{n})\pi_{n}(k)-\beta_{n}\pi_{n-1}(k)
\end{eqnarray}
with $\pi_{-1}(k)=0$, $\pi_{0}(k)=1$, and $n=0,\cdots,M-1$. The coefficients of this recurrence, $\alpha_{n}$ and $\beta_{n}$, can be obtained with standard numerical routines \cite{gautschi2005}. 
Hence, the resulting Hamiltonian describes two tight-binding chains to which the system is coupled. The thermofield vacuum 
is also annihilated by the new modes $B_n$, so that the dynamics of the whole system can be simulated 
using MPS time-evolution methods from an initial state with zero occupancy of each of these modes.
Note that a similar mapping can be applied in the case of 
a finite discrete environment by means of 
a standard Lanczos tri-diagonalization.

%
%
\bibliography{Bibtexelesdrop_NMloc2}
\bibliographystyle{prsty}

\end{document}